\documentclass[aps,prl,reprint,superscriptaddress]{revtex4-1}
\usepackage{graphicx}% Include figure files
\usepackage{dcolumn}% Align table columns on decimal point
\usepackage{epstopdf}
\usepackage{rotating}
\usepackage{amssymb}
\usepackage{mathptmx}
\usepackage{bm}

\begin{document}

\title{Orbitropic Effect in Superfluid $^3$He B-phase boundaries}


\author{M. Array\'as}
\affiliation{\'Area de Electromagnetismo, Universidad Rey Juan Carlos, Tulip\'
an s/n, 28933 M\'ostoles, Madrid, Spain.}
%\thanks{}

\author{R. Haley}
\author{G. R. Pickett}
\author{D. E. Zmeev}
%\thanks{}
\affiliation{Department of Physics, Lancaster University, Lancaster LA1 4YB, UK.}


\date{\today}

\begin{abstract}
In this work we study the influence of orbital viscosity on the dynamics of the order parameter texture in the superfluid B-phase of $^3$He near a moving boundary. Based on the redistribution of thermal quasiparticles within the texture, we develop a model which bestows a significant effective mass on the interface, and gives a new mechanism for friction as the boundary moves. We have tested the model against existing data of a moving A-B interface whose motion was controlled using magnetic field. The model allows some predictions in experimental situations which involve texture rearrangement due to the B-phase boundary motion. 

\end{abstract}

%\keywords{}

\maketitle

%\section{The theory}

Many systems \cite{Osheroff:72, Anderson:95, Davis:95, Kasprzak:06, Nikuni:00, Bunkov:07, Page:11} condense into coherent states characterised by a ``rigidity'' to distortions in their wave function description \cite{London:50}. Among such systems, superfluid $^3$He at very low temperature can condensate in two phases A and B. The bulk properties of the A- and B- phases are presently very well understood \cite{Tilley}. However close to the boundaries, the topology of the order parameter changes and one might expect unconventional behaviour \cite{Leggett:90}. The work described here is motivated by the unexpected dissipation observed in actual experiments involving B-phase boundaries.
We have found a new mechanism for friction inside a superfluid condensate that is generally thought to be frictionless. This mechanism should be relevant to all multiple-phase coherent condensates with non-trivial topologies. We show, the change in the topology of the orbital angular momentum has associated with it an effective inertia and damping. We are able to make quantitative predictions and successfully test these against previously unexplained observations, as well as propose future experiments.
  
Superfluid $^3$He can be thought of as comprising three separate components: the mass superfluid, the spin superfluid and the orbital superfluid. The dynamics of the orbital superfluid is so strongly clamped by the normal fluid that it has been ignored for the whole history of superfluid $^3$He. In zero magnetic field the superfluid B-phase of $^3$He is pseudo-isotropic, with no net spin or orbital angular momentum and an isotropic energy gap $\Delta_0$ \cite{Vollhardt:90}. However, in a magnetic field the B-phase order parameter is distorted and the energy gap acquires a minimum along the orbital anisotropy axis $\boldsymbol{l}_B$. When the magnetic field is strong (like the one needed to stabilize the A-phase), the B-phase anisotropy is dominated by Zeeman splitting thus a large density of thermal excitation occupy states along the $\boldsymbol{l}_B$ axis. Any sudden change in the spatial orientation of $\boldsymbol{l}_B$ will generate substantial dissipation as the thermal excitations must redistribute. This redistribution has both a dissipative component which can be related to orbital viscosity \cite{Fisher:05}, and a reactive one which can be characterized as an effective mass \cite{Arrayas:2014}. 

Usually, the orbital textures bend near the boundaries. For example, in cylindrical geometries, with the magnetic field parallel to the cylindrical axis, the energetically favorable spatial distribution of $\boldsymbol{l}_B$ is the ``flare-out'' texture \cite{Brinkman:74,Smith:77,Hakonen:89}. Here,  $\boldsymbol{l}_B$ is  oriented parallel to the magnetic field far from the walls and bends so it comes perpendicular to the side walls.  Another place to look at the change of the orbital axis $\boldsymbol{l}_B$ is at the  interface between the A-phase and the B-phase, stabilized by a magnetic field gradient (at low temperatures and pressures the critical field for the A-B phase transition is $B_c=$340\,mT). In the bulk of the B-phase $\boldsymbol{l}_B$ points along the magnetic field direction, but closer to the interface it is energetically favorable for it to be parallel to the interface \cite{Thuneberg:91}. 

In this work we study the effect of the change of the orbital direction at the boundaries of the B-phase in magnetic fields due to the movement of these boundaries. As we are interested in the change of the tilt angle, taking as reference the direction of the magnetic field, this can be considered a nutation motion when described using Euler angles. We apply the results  to describe the dynamical behavior of an oscillating A-B interface in high magnetic fields at very low temperatures and to give some predictions when the B-phase boundaries move. The dissipation will depend on the orientation of the movement of the boundary with respect to the equilibrium texture configuration.    

In a magnetic field the energy of a quasiparticle excitation with momentum $p$ and spin $\sigma\hbar$ is given by \cite{Schopohl:82,Ashida:85}
\begin{equation}
E_{p,\sigma} = \sqrt{(E_{\Vert}({\bf p}) - \sigma \hbar\widetilde{\omega}_L)^2 + (\Delta_{\bot}p_{\bot})^2}
\end{equation}
where $E_{\Vert}({\bf p}) = (\xi^2 + \Delta_{\Vert}^2p_{\Vert}^2)^{\frac{1}{2}}$ and $\xi = (p-p_F)v_F$ is the kinetic energy relative to the Fermi energy. Here $\Delta_{\Vert}$ and $\Delta_{\bot}$ are the energy gaps parallel and perpendicular to ${\boldsymbol l}_B$, $p_{\Vert}$ and $p_{\bot}$ are the parallel and perpendicular components of the quasiparticle momentum, $p_F$ is the Fermi momentum and $v_F$ is the Fermi velocity. The Fermi-liquid corrected quasiparticle Zeeman energy is $\sigma \hbar\widetilde{\omega}_L$ with $\sigma = \pm 1/2$. In practice, the Zeeman splitting dominates the anisotropy and so $\Delta_{\Vert}$ and $\Delta_{\bot}$ can be approximated by the zero field gap $\Delta_0$~\cite{Ashida:85,Nagai,FisherThesis}.  The minimum quasiparticle energy is a function of the angle between its momentum and the ${\boldsymbol l}_B$ axis. At temperatures far below $T_c$ the vast majority of excitations occupy the lowest energy states with momenta centered around the ${\boldsymbol   l}_B$ axis. Any distortion of the local orientation of ${\boldsymbol l}_B$ will change the quasiparticle energies. The subsequent relaxation of the excitations occurs over a time scale $\tau$. This is the essential mechanism for providing the effective mass and friction of the boundaries under non-stationary conditions. 

Taking  $\theta$ to be the angle between the magnetic field and the ${\boldsymbol l}_B$ vector, a change in the orientation of ${\boldsymbol l}_B$ by an amount $\delta \boldsymbol{\theta}$ changes the quasiparticle energies by
\begin{equation}
\delta E_{p,\sigma} =
 \frac {\partial E_{p,\sigma}} {\partial \boldsymbol{\theta}}\cdot { \delta \boldsymbol{\theta} }
\label{eq:deltaE}
\end{equation}
which produces a viscous torque \cite{Cross:77, Brinkman:78}
\begin{equation}
{\boldsymbol \Gamma_{vis}} = - \mu \boldsymbol{l}_B \times \dot{\boldsymbol {l}}_B = -\mu \dot{\boldsymbol{\theta}},
\end{equation}
where $\mu$ is the orbital viscosity. The orbital viscosity $\mu$ in the B-phase at low temperatures was previously shown to be \cite{Fisher:05}
\begin{equation}
\mu = \frac{\tau + i \omega \tau^2}{1+(\omega\tau)^2} \left[\frac{\pi}{6}N(0)\frac{\Delta}{k_BT}\exp{(-\Delta/k_BT)}(\hbar \widetilde{\omega}_L)^2\right]
\label{mulow}
\end{equation}
to first order in $\widetilde{\omega}_L^2$, 
where $\tau$ is quasiparticle relaxation time, 
$N(0)$ is the normal density of states at the Fermi surface, and $\Delta$ is approximately equal to the zero field gap $\Delta_0$.

Let us assume that the orbital texture in the B-phase responds instantaneously to the changing position of the interface. This means that we neglect the orbital dynamics and suppose that $\boldsymbol{l}_B$ always has its equilibrium orientation $\boldsymbol {\theta} ({\bf r} - {\bf r}_I )$ relative to the position of the boundary interface ${\bf r}_I$. This adiabatic approximation is
ultimately justified as long as the characteristic time of the motion of the boundary is much greater than $\tau$. A fuller treatment in the future ought to take into account the dynamics of the $\boldsymbol{l}_B$ texture itself.
Now consider a small change in the position of the boundary interface $\delta {\bf r}_I$. This causes the texture orientation in the B-phase to adjust by
\begin{equation}
\delta \boldsymbol {\theta} ({\bf r}) = \nabla \boldsymbol {\theta} \cdot \delta {\bf r}_I
\label{eq:deltatheta}
\end{equation}
and thus
\begin{equation}
\dot{\boldsymbol{\theta}} = \nabla \boldsymbol {\theta} \cdot \dot{{\bf r}}_I.
\label{eq:thetadot}
\end{equation}
Note that  $\nabla{\boldsymbol \theta}$ is a tensor of the second order.  The corresponding work done on the quasiparticle distribution  is
\begin{equation}
\delta W = -\int_{V}\boldsymbol {\Gamma}_{vis}\cdot \delta\boldsymbol {\theta} dV = \int_{V}\mu \dot {\boldsymbol {\theta}}\cdot \delta\boldsymbol {\theta} dV
\label{thetawork}
\end{equation}
where the integral is over the whole volume of the B-phase. 

We can find the corresponding effect on the boundary dynamics by considering the work done by the viscous torque (\ref{thetawork}) and equating it to the work done by a moving boundary. The force exerted per unit surface by the moving boundary, considering only the effect of the viscous torque, will have a reactive and dissipative components corresponding to the inertial and frictional parts respectively ${\bf F}=m_l \ddot{\bf r}_I + \gamma_l \dot{\bf r}_I$. We can decompose the motion into the Fourier modes, so $\ddot{\bf r}_I = i \omega \dot{\bf r}_I$ and the work done on moving the interface by $\delta {\bf r}_I$ can be written as
\begin{equation}
\delta W = \int_{s}(\gamma_l + i \omega m_l ) \dot{{\bf r}}_I\cdot \delta {\bf r}_I\,dS.
\label{eq:intwork}
\end{equation}
Equating this with the work done on the quasiparticles, Eq.~(\ref{thetawork}), and using expressions (\ref{eq:deltatheta}) and (\ref{eq:thetadot}) gives
\begin{equation}
\int_{s}(\gamma_l + i \omega m_l)\dot{{\bf r}}_I\cdot \delta {\bf r}_I dS = \mu \int_{V}(\nabla \boldsymbol {\theta} \cdot \dot{{\bf r}}_I)\cdot (\nabla \boldsymbol {\theta}\cdot   \delta {\bf r}_I)\,dV.
\label{mainresult}
\end{equation}
Thus the friction coefficient $\gamma_l$ is related to the real part of the orbital viscosity $\mu$ and the effective mass of the interface $m_l$ is related to the imaginary part.

\begin{figure}
    \begin{center}
       \includegraphics[width=0.95\linewidth]{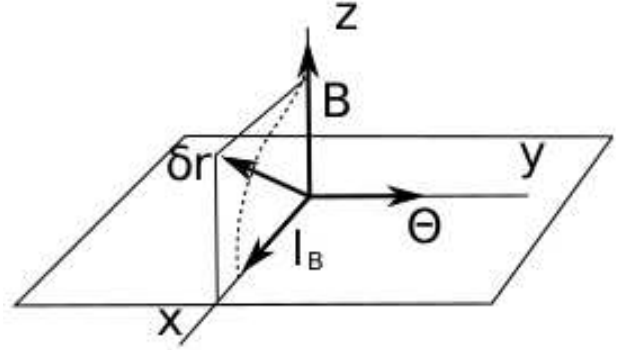}
       \caption{Sketch of the case discussed in the main text. The interface lies in the $xy$ plane. In the bulk the magnetic field ${\bf B}$ points to the $\boldsymbol{z}$ direction so does the $\boldsymbol{l}_B$ texture. On approaching to the interface $\boldsymbol{l}_B$ must rotate to become parallel to the interface.}
        \label{fig:Fig1s}
    \end{center}
\end{figure}

\begin{figure}
    \begin{center}
       \includegraphics[width=0.95\linewidth]{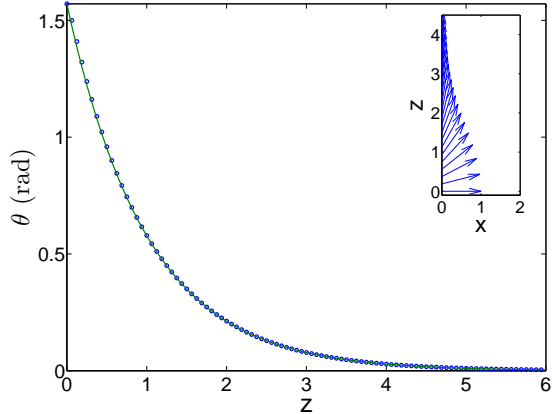}
       \caption{Numerical calculation of the $\boldsymbol{l}_B$ texture near the A-B interface. $\theta$ is the angle between $\boldsymbol{l}_B$ and the $\boldsymbol{z}$ axis. The solid line is the plot of $({\pi}/{2}) \exp{(-(z-z_I)/\xi_B)}$ taking $z_I$ at the origin. In the inset, it is plotted $\boldsymbol{l}_b$ as a vector of unit length in arbitrary units, showing the change in the orientation. The $\boldsymbol{x}$ and $\boldsymbol{z}$ axes are in  units of $\xi_B$.}
        \label{fig:Fig1}
    \end{center}
\end{figure}

Equation~(\ref{mainresult}) allows to make some predictions. Let us take the simplest case where the equilibrium texture configuration is parallel to the interface which is in the horizontal direction, and in the bulk the $\boldsymbol{l}_B$ texture is vertical, in the direction of the magnetic field $\boldsymbol{B}$. The situation is shown in Fig.~\ref{fig:Fig1s}. In this configuration  $\boldsymbol{l}_B$ must rotate to become horizontal on approach to the interface \cite{Thuneberg:91}. The change in orientation occurs over a distance of order $\xi_B$ . This texture has been calculated numerically by finding the configuration which minimizes the bending energy and the magnetic free energy simultaneously \cite{Brinkman:78}.  It is plotted in Fig.~\ref{fig:Fig1} where $\theta$ denotes the angle between ${\boldsymbol l}_B$ and the vertical axis $\boldsymbol{z}$. 

Thus, in this particular situation the texture ${\boldsymbol\theta}(z)$ is  described well by
\begin{eqnarray}
  \label{eq:a11}
  \theta_x&=&\theta_z=0,\\
  \theta_y&=&\frac{\pi}{2}\exp{(-(z-z_I)/\xi_B)}\label{eq:a1}.
\end{eqnarray}
Now take as the displacement of the interface a translation in the $\boldsymbol{x}$ and $\boldsymbol{z}$ directions,
\begin{equation}
  \label{eq:a2}
  \delta {\bf r}_I = \delta r (\cos\phi,0,\sin\phi),
\end{equation}
so
\begin{equation}
  \label{eq:a3}
  \dot{{\bf r}}_I = \dot{r} (\cos\phi,0,\sin\phi).
\end{equation}

Then using (\ref{mainresult}) and (\ref{eq:a1}) we get 
\begin{equation}
\label{eq:a4}
\int_{s}(\gamma_l + i \omega m_l) dS = \mu \int_{V}\left(\frac{\partial \theta_y}{\partial z_I}\right)^2\sin^2\phi\,dV.
\end{equation}

Thus, for $\phi=0$, which corresponds to a boundary displacement parallel to the interface, the effective mass and dissipation disappears, while the maximum will occur for a displacement in the normal direction. It must be noted that the presence of any defect in the texture on the boundary  or a different equilibrium configuration of the texture will change this prediction, but nevertheless we expect that there should be preferred directions for the effective mass and the dissipation.  

We can apply our model to an oscillating A-B interface. The dynamics of the A-B interface itself is particularly interesting at low temperatures.  It has been proposed that the interface has an effective mass and there is dissipation due to pair-breaking and Andreev scattering when it moves \cite{Leggett:90}. Indeed at higher interface velocities the pair-breaking has analogies with Schwinger pair-creation and the Unruh effect in particle physics~\cite{Schopohl:92}. It was found experimentally \cite{Buchanan:1986gu} that the friction values of a fast freely-moving interface in low magnetic fields were in line with theoretical estimates based on the Andreev scattering of thermal quasiparticle excitations \cite{Leggett:86}. Later measurements made in Lancaster in high magnetic fields and much lower temperatures \cite{Bartkowiak:2000tja} found the friction to be orders of magnitude higher than the theoretical predictions \cite{Leggett:90,Leggett:86,Kopnin:87}.  The measurements were made on an interface that was stabilized and driven into controlled oscillation using shaped magnetic field profiles, and at much lower temperatures where pair-breaking was expected to dominate the dissipation~\cite{Bartkowiak:2000wh}. Furthermore, the dissipation showed non-linear behavior that appeared to depend on the frequency of the oscillations of the interface. It was pointed that the dynamics of an oscillating A-B interface in high magnetic fields and low temperatures is dominated by orbital viscosity and by a significant effective mass generated by thermal quasiparticle excitations in the B-phase order parameter texture~\cite{Arrayas:2014}. The change of the surrounding texture and energies of thermal quasiparticles should contribute to the effective mass of the interface and also produce dissipation due to orbital viscosity~\cite{Fisher:05}. 

The motion of the interface can be described by \cite{Leggett:86}
\begin{equation}
m \ddot{{\bf r}}_I + \gamma \dot{{\bf r}}_I = {\bf n}_{AB}\Delta G_{AB},
\label{eq:motion1}
\end{equation}
where ${\bf n}_{AB}$ the surface normal directed toward the B-phase, and $\Delta G_{AB} = 1/2 \chi_{AB}(B_c^2-B^2)$ the Gibbs energy difference per unit volume for the two phases with $\chi_{AB}$ being the difference in the magnetic susceptibilities of the A- and B- phases.

In order to simplify the calculations, we will further assume as before that the B-phase responds instantaneously to the changing position of the interface and that the boundary remains flat during its motion. This last approximation is valid as long as the healing length $\xi_B$ over which ${\boldsymbol   l}_B$ changes direction is smaller than the size of the boundary and the effect of the side walls in the form of a meniscus can be neglected. We note, that in the A-phase the preferred orientations of the orbital vector ${\boldsymbol   l}_A$  in the bulk and on the surface are the same, so the orbitropic effect does not manifest in the A-phase when the interface is moving along the direction of the magnetic field.

Under those conditions Eq.\,(\ref{mainresult}), after substituting $dV = A dz$ with $A$ the area of the interface, reads
\begin{equation}
\gamma_l + i \omega m_l = \mu \int_{z} \left(\frac{\partial \theta_y}{\partial z_I}\right)^2 dz.
\label{gammavis1}
\end{equation}
We can estimate the integral in Eq.\,(\ref{gammavis1}) by supposing that $\theta$ changes exponentially from $\frac{\pi}{2}$ to zero over the textural healing length $\xi_B$, as given by Eq.\,(\ref{eq:a1}). The integral in Eq.\,(\ref{gammavis1}) is then evaluated as
\begin{equation}
\int_{z}\left(\frac{\partial \theta_y}{\partial z_I}\right)^2 dz = \frac{\pi^2}{8\xi_B}=\frac{1.2337}{\xi_B}.
\label{intcalc}
\end{equation}
The numerical evaluation of this integral based on the texture configuration which gives the minimum for the bending energy and the magnetic free energy simultaneously results $1.2353/\xi_B$ instead. 

Substituting Eqs.\,(\ref{mulow}) and (\ref{intcalc}) in Eq.\,(\ref{gammavis1}) gives the following estimates for the friction   coefficient 
\begin{equation}
\gamma_l = \frac{\tau}{1+(\omega\tau)^2} \left[\frac{\pi}{6}N(0)\frac{\Delta}{k_BT}\exp{(-\Delta/k_BT)}(\hbar \widetilde{\omega}_L)^2\right]  \frac{\pi^2}{8\xi_B}
\label{gammavis2}
\end{equation}
and the effective mass
\begin{equation}
m_l = \frac{\tau^2}{1+(\omega\tau)^2} \left[\frac{\pi}{6}N(0)\frac{\Delta}{k_BT}\exp{(-\Delta/k_BT)}(\hbar \widetilde{\omega}_L)^2\right] \frac{\pi^2}{8\xi_B}.
\label{mass2}
\end{equation}
So that the friction and mass parameters of Eq.\,(\ref{eq:motion1}) become $m=m_l$ and $\gamma=\gamma_0+\gamma_l$. The contribution  $\gamma_0$ comes from Andreev scattering and pair breaking~\cite{Leggett:90,Leggett:86,Kopnin:87}, or any dynamical friction with the walls.
  
Let us test the results on the experimental data . The experiments were performed on a sample of superfluid $^3$He contained in a sapphire tube connected to the inner cell of a Lancaster-style nuclear cooling stage. They are described earlier in more detail \cite{Bartkowiak:2000tja}. The sapphire tube has internal diameter 4.3\,mm and length 44\,mm. The experiments were carried out at a pressure of 0\,bar . A solenoid stack was used to create a shaped magnetic field profile to stabilize the A-phase in the bottom of the tube in fields above the transition field $B_c\,=\,340\,{\rm mT}$~\cite{Bartkowiak:2000tja, Bartkowiak:1999wy, Hahn:94}, whilst maintaining the top of the tube in the low field B-phase. Once the A-B interface was established across the tube, a small additional alternating field was applied to oscillate the interface over a range of frequencies.

\begin{figure}
    \begin{center}
       \includegraphics[width=1.0\linewidth,keepaspectratio]{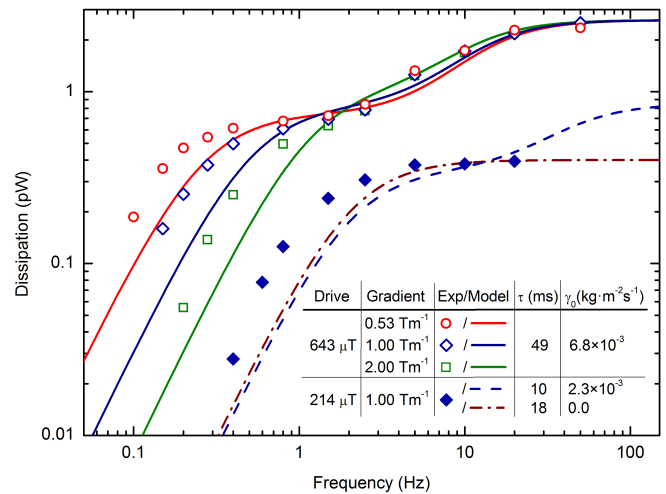}
        \caption{(color online)  Dissipation of the oscillating A-B interface versus frequency. The points show measurements at a field oscillation amplitude of $B_{ac}=0.643$\,mT for  three different field gradients 2.00, 1.00 and 0.53\,T/m, and $B_{ac}=0.214$\,mT for the gradient 1.00\,T/m. The lines are fits to the model (see text). The inset table summarizes the results of the fits.}
        \label{datafit}
    \end{center}
\end{figure}

To evaluate $\gamma_l$ and $m_l$ for the experimental data in Fig.~\ref{datafit} we took the following values: $\hbar \widetilde{\omega}_L = 0.67\Delta_0$, from~\cite{Ashida:85,Nagai,FisherThesis}; $\Delta_0 = 1.76 k_B T_c$ with $T_c = 0.929$\,mK at 0\,bar pressure, from~\cite{Greywall:86}; $N(0) = 1.0 \times 10^{51}$\,m$^{-3}$, from~\cite{Greywall:86,Wheatley:75}; and $T=155$\,$\mu$K (noting that this is an average temperature, and the actual temperature during the measurements varied from $T=150$\,$\mu$K to $T=160$\,$\mu$K as the dissipation increased from low to high frequencies).  We assume $\xi_B=0.1$ mm, using some estimations for fields close to $B_c$ \cite{Alles:99,Thuneberg:91,Hakonen:89,Ishikawa:89,Kopu}.

The only unknown parameters in our model are the quasiparticle relaxation time $\tau$ and the frequency independent dissipation term $\gamma_0$, which is an  additive term to  $\gamma_l$ in (\ref{gammavis2}). The model predictions are shown together with the experimental data in Fig.~\ref{datafit}. We have found that two values of $\tau$ on the order of tens of milliseconds are needed to fit the experimental data, depending on the magnitude of the driving magnetic field. The value for $\tau$ calculated for the {\it uniform} texture and for the given experimental conditions is only a few milliseconds \cite{Einzel:78}. With the texture bending, the quasiparticles may become trapped within the texture, which would increase their relaxation time.  In the case of the dissipation parameter $\gamma_0$,  two different values are needed as well to successfully fit the measured dependence. The values of $\gamma_0$ roughly scale as the amplitudes of the driving field (their ratio being about 3). For the set of experimental data in the case of the $214\,\mu$T drive, we have also considered the fitting with no extra constant dissipation. The lack of experimental data at higher frequencies prevents us from drawing further conclusions. We remark that texture bending could be a function of the driving amplitude and remains to be further investigated.

In future experiments the orbitropic effect can be studied in other, less complicated situations. For example, in a magnetic field in the vicinity of a thin  superconducting wire carrying  alternating current. The magnetic field generated by this current will reorient the texture in $^3$He-B, thus causing dissipation (the magnetic field on the surface of a  0.1\,mm thick wire carrying 1\,A is 4\,mT and the field gradient is 80\,T/m).

Finally, we would like to comment on the results of a recent paper concerning the motion of the A-B interface \cite{Todo:16}. The paper predicts a very peculiar form of a magnetic wave that can be excited at the interface: as the two superfluid phases have different magnetic susceptibilities, the phase transition between them is accompanied by a change in magnetisation. The author estimates that the effective inertia of such a wave  is not large enough  for the wave to be excited in a typical experiment. However, the orbitropic effect will endow the wave with a high value of the effective mass: at the resonant frequency of the wave ($f_\mathrm{res}=220$\,Hz for typical experimental conditions) our calculation (\ref{mass2}) gives $m_l = 8.5 \times 10^{-7}$\,kg/m$^2$ which is about 100 times greater than that predicted in  \cite{Todo:16}. As a result, the dissipation rate of 35\,s$^{-1}$ is small compared to  $2\pi f_{\mathrm{res}}$ and makes the observation of the magnetic phase wave more realistic.

In conclusion, in this work we have considered  a new dissipation mechanism based on the redistribution of thermal quasiparticles in the orbital texture.  We have investigated its  effect at the boundaries of the B-phase, and studied the consequences of the change of the nutation angle of the orbital vector $\boldsymbol{l}_B$. We have estimated the effective mass and dissipation associated with this motion, assuming adiabatic approximation of the influence of the orbital dynamics on the boundary dynamics. The main result is expressed in Eq.~(\ref{mainresult}). Further, we have applied the theory developed to the dynamics of an oscillating A-B interface. The results are in a reasonable agreement with the experimental data  and explain the increase in dissipation for frequencies above 2 Hz which could not be explained by existing theories.    

The research is supported by the UK EPSRC and by the European FP7 Programme MICROKELVIN, Project no 228464. M.~Array\'as is also supported by Spanish Ministerio de Econom\'{\i}a y Competitividad, under project ESP2015-69909-C5-4-R. We acknowledge the late S. N. Fisher for his contribution to the early stages of this work. 

\bibliographystyle{apsrev4-1}
\bibliography{biblioHe}

\end{document}